# All-optical nonreciprocity due to valley polarization in transition metal dichalcogenides

Yuma Kawaguchi[1*], Sriram Guddala[1*], Kai Chen[1], Andrea Alù[3,2,1], Vinod Menon[2,4], and Alexander B. Khanikaev[1,2,4]

[1]Department of Electrical Engineering, Grove School of Engineering, City College of the City University of New York, 140th Street and Convent Avenue, New York, NY 10031, USA.
[2]Physics Program, Graduate Center of the City University of New York, New York, NY 10016, USA.
[3]Photonics Initiative, Advanced Science Research Center, City University of New York, New York, NY 10031, USA
[4]Department of Physics, City College of New York, 160 Convent Ave., New York, NY 10031, USA

**Abstract**
Nonreciprocity and nonreciprocal optical devices play a vital role in modern photonic technologies by enforcing one-way propagation of light. Most nonreciprocal devices today are made from a special class of low-loss ferrites that exhibit a magneto-optical response in the presence of an external static magnetic field. While breaking transmission symmetry, ferrites fail to satisfy the need for miniaturization of photonic circuitry due to weak character of nonreciprocal responses at optical wavelengths and are not easy to integrate into on-chip photonic systems. These challenges led to the emergence of magnetic-free approaches relying on breaking time reversal symmetry, e.g. with nonlinear effects modulating optical system in time. [1–5] Here, we demonstrate an all-optical approach to nonreciprocity based on nonlinear valley-selective response in transition metal dichalcogenides (TMDs). This approach overcomes the limitations of magnetic materials and it does not require an external magnetic field. We provide experimental evidence of photoinduced nonreciprocity in a monolayer $WS_2$ pumped by circularly polarized light. Nonreciprocity stems from valley-selective exciton-exciton interactions, giving rise to nonlinear circular dichroism controlled by circularly polarized pump fields. Our experimental results reveal a significant effect even at room temperature, despite considerable intervalley-scattering, showing potential for practical applications in magnetic-free nonreciprocal platforms. As an example, we propose a device scheme to realize an optical isolator based on a pass-through silicon nitride (SiN) ring resonator integrating the optically biased TMD monolayer.

**Introduction**
Nonreciprocal optical devices, such as isolators and circulators, are critical components for photonic systems[3–12] at large. Optical isolators enable stable laser operation by blocking reflected light from entering the laser cavity, and circulators facilitate nonreciprocal routing of optical signals in telecommunication networks. However, nonreciprocal devices available today rely on magneto-optical materials, which have limited possibility of integration into modern photonic circuitry due to chemical incompatibility of materials and typically require a bulky external magnetic bias. In addition, the weak character of magneto-optical effects prevents miniaturization of magneto-optical components, which must be large to provide sufficient nonreciprocal response. Although numerous solutions have been proposed, including photonic[1,5,13–16] and plasmonic materials[17–20] integrating magneto-optical media, these schemes have not yet been proven to be of technological relevance.

In recent years, magnet-free approaches to nonreciprocity have gained attention, including linear[1]- and angular-[2] momentum biased photonic structures and metamaterials. In such systems, parametric phenomena induced by external time-modulated bias were shown to give rise to nonreciprocal responses. However, electro-optical modulation schemes are limited to a few GHz speeds, implying that optical non-reciprocity can be difficultly achieved with these schemes, and most experimental demonstrations with practically relevant metrics of performance have been limited to radio-frequencies[3,22]. Nonlinear phenomena combined with asymmetric field distributions have also been shown to enable nonreciprocity in some regimes[23], exploiting the temporal modulations enabled by the signal itself as it propagates through the device[24,25]. However, this form of self-bias nonreciprocity comes with some drawbacks [22,26–28] that hinder its widespread applicability. In the optical domain, therefore, magnet-free isolators and circulators remain elusive, even though some important proof-of-concept experimental schemes have been demonstrated[4,5]. Nonetheless, some of the all-optical modulation schemes to break reciprocity proposed recently may overcome the above limitations[29].

In a different context, two-dimensional (2D) Van-der-Waals materials have been shown to provide a promising platform for enhanced light-matter interactions, including enhanced nonlinear responses in graphene and other 2D materials[30–35]. A particular class of 2D semiconductors, monolayer transition metal dichalcogenides (TMDs), has attracted significant attention from the research community due to their unique valley-dependent optical response[36–39]. The conservation of angular momentum in TMDs enforces circularly polarized light to interact selectively with electronic subsystems at K and K' valleys, leading to valley-selective absorption of circularly polarized light[40–43]. Valley-polarized excitons have been shown to support circularly polarized luminescence connected with the pump handedness, due to the conservation of the valley degree of freedom. A variety of fascinating effects based on such chiral light-mater interactions have been demonstrated recently, including directional launching of guided waves and surface plasmon-polaritons[42,44–46]. More recently nonlinear effects in TMDs such as saturable absorption[47], valley-dependent exciton bistability[48] and valley-dependent second harmonic generation[49–51] have been demonstrated and proposed for valley optoelectronic applications.

In this work, we exploit the enhanced nonlinearity in TMDs, and the valley selectivity in their nonlinear response, to experimentally demonstrate that nonlinear chiral light-matter interaction in TMDs open a route to all-optical nonreciprocal photonics. We show that valley-selective exciton-exciton interactions lead to photoinduced nonreciprocal circular dichroism analogous to the one observed in magneto-optical materials, but in which optical pumping with given handedness replaces the magnetic bias.

**Photoinduced nonreciprocal circular dichroism**

The scheme illustrating the concept of photoinduced nonreciprocity is shown in Fig. 1**a**. A TMD monolayer is pumped by a strong circularly polarized laser radiation, which leads to the selective formation of exciton gas at one of the valleys. At high pump field intensities, the increasing density of excitons at the photoexcited valley gives rise to stronger nonlinear response due to the exciton-exciton interactions[52] (Fig. 1b). Provided the valley polarization of excitons is at least partially preserved, it will give rise to an asymmetric response of the TMD monolayer to weak probe signals of opposite handedness, due to the fact that they selectively interact with one of the two valleys[44]. As schematically depicted in Fig. 1a, this effect leads to a nonreciprocal dichroic response, i.e., probe signals of opposite handedness are absorbed and reflected differently from the optically pumped TMD monolayer. Since the handedness of circularly polarized light is locked to the propagation direction, and similar locking of transverse angular-momentum to propagation direction exists for evanescent electromagnetic fields[53–55], this opens the opportunity for designing optical elements with inherently nonreciprocal response induced by the circularly polarized pump field. Is worth noting here that the circular dichroic response of any planar 2D systems, including 2D materials, is necessarily *nonreciprocal*, since *reciprocal* circular dichroism, known as optical activity, requires nonlocal bianisotropic response, which is possible only in structures with finite thickness[56,57].

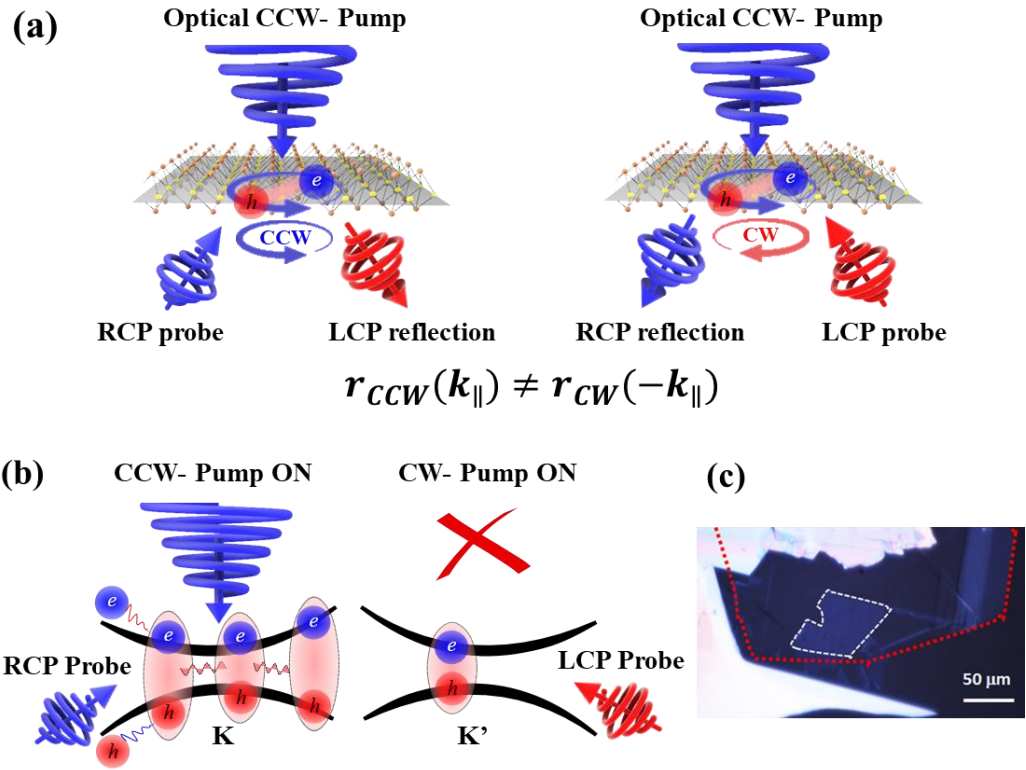

**Figure 1| Photoinduced nonreciprocity in TMDs. a**, Schematic illustration of nonreciprocal reflection due to valley-selective response induced by a circularly polarized pump. **b**, Mechanisms for long-living dichroic response and nonlinear saturation as a two-step process with delayed relaxation of free-carriers to excitons and of exciton-exciton interactions. **c**, Optical microscope image of encapsulated $WS_2$ monolayer on glass substrate. Encircled white and red dotted lines refer to the WS2 monolayer and the thin layer of hBN respectively.

First, we investigate the optical response of a $WS_2$ monolayer under circularly polarized pump, expecting to observe nonreciprocal circular dichroism at large pump intensities. As schematically shown in Fig. 1**a**, a pump with left-handed circular polarization illuminating from above the sample has a counter clock-wise (CCW) projected helicity on the plane of the 2D material, which leads to the formation of excitons at the K valley in WS$_2$. In the ideal case of no valley scattering, this increased exciton density at one of the valleys leads to exciton-exciton interactions (schematically shown in Fig. 1**b**) which affects reflectivity of a probe signal of the same helicity as the pump, while the reflectivity of the probe field with opposite helicity remains unaffected. As a consequence, probe signals illuminating from opposite angles with opposite helicity, time-reversed versions of each other, will experience different reflectivity $r_{CW} \neq r_{CCW}$. Since the projected helicity is flipped for opposite propagation directions of the probe signal, as illustrated in Fig. 1**a**, this response is nonreciprocal $r_{CW}(k_\parallel) \neq r_{CCW}(-k_\parallel)$, and can be used to realize an all-optical magnet-free isolator, as shown in the following.

**Theoretical model**

In order to quantitatively describe the dichroic nonreciprocal response in optically pumped TMDs, we introduce a nonlinear surface conductivity tensor

$$\hat{\sigma} = \sigma_N(I) + \hat{\sigma}_K(I_{CW}) + \hat{\sigma}_{K'}(I_{CCW}), \tag{1}$$

where $\sigma_N$ describes the valley-independent optical response, $\hat{\sigma}_K$, $\hat{\sigma}_{K'}$ correspond to the valley-dependent response due to excitations at K and K' valley, respectively, $I_{CW}$, $I_{CCW}$ are the intensities of the circularly polarized pump fields, which make clockwise and counterclockwise projections of the electric field onto the TMD plane, respectively, and $I = I_{CW} + I_{CCW}$ is the total pump field. In addition to conventional non-valley polarized optical processes, the first term in Eq. (1), $\sigma_N(I)$, also accounts for valley "depolarization" due to various inter-valley scattering processes. It is worth highlighting that here the notations of CW and CCW specify the handedness of the electric field rotation in the TMD plane, irrelevant of the propagation direction, and that LCP and RCP polarization/handedness of optical waves are therefore not in one-to-one correspondence with CW and CCW. The valley-polarized response is uniquely described by the projected helicity (CW/CCW).

The valley-dependent terms in (1) have the following form, which accounts for their chiral response:

$$\hat{\sigma}_K = \frac{1}{2}\begin{pmatrix} \sigma_K & i\sigma_K \\ -i\sigma_K & \sigma_K \end{pmatrix}, \hat{\sigma}_{K'} = \frac{1}{2}\begin{pmatrix} \sigma_{K'} & -i\sigma_{K'} \\ i\sigma_{K'} & \sigma_{K'} \end{pmatrix}, \tag{2}$$

where $\sigma_K = \sigma_K(I_{CW})$ and $\sigma_{K'} = \sigma_{K'}(I_{CCW})$ are surface conductivities for the two valleys in the circularly polarized (CP) basis. The form of Eqs. (2) follows directly from the fact that the response of each valley is selective with respect to the handedness of the optical field, and therefore it is described by the matrices $\hat{\sigma}_K^{CPB} = [\sigma_K, 0; 0, 0]$ and $\hat{\sigma}_{K'}^{CPB} = [0, 0; 0, \sigma_{K'}]$ in the circularly polarized basis (CBP).

In the case of no optical pump, the two valleys yield the same response $\sigma_K(I_{CW} = 0) \equiv \sigma_{K'}(I_{CCW} = 0)$, so that $\hat{\sigma}_K + \hat{\sigma}_{K'} = [\sigma_K, 0; 0, \sigma_{K'}]$ and the TMD shows no asymmetry in the response with respect to CW and CCW probe signals. However, dichroism arises as the pump intensity of a particular handedness is increased, and the response enters the nonlinear regime such that $\sigma_K(I_{CW}) - \sigma_{K'}(I_{CCW}) \neq 0$, yielding an effective response of the form

$$\hat{\sigma}_{TMD} = \begin{pmatrix} \sigma_N + \frac{1}{2}(\sigma_K + \sigma_{K'}) & \frac{i}{2}(\sigma_K - \sigma_{K'}) \\ -\frac{i}{2}(\sigma_K - \sigma_{K'}) & \sigma_N + \frac{1}{2}(\sigma_K + \sigma_{K'}) \end{pmatrix} = \begin{pmatrix} \sigma_{xx} & i\sigma_{xy} \\ -i\sigma_{xy} & \sigma_{xx} \end{pmatrix}. \quad (3)$$

This response is equivalent to the one of a 2D electron gas in the presence of a dc external magnetic bias[58,59], showing how the circularly polarized optical pump can effectively break time-reversal symmetry in TMDs. Indeed, the possibility to use a circularly polarized pump as an effective magnetic field bias has been proposed before in the context of so called Floquet systems and Floquet topological insulators have been introduced[29,60,61]. More recently, the circularly polarized pump field was used to demonstrate photoinduced quantum Hall effect in graphene[62]. Here, however, we report the effect not related to Floquet physics, but originating solely in the valley-polarized nonlinear optical response of TMDs.

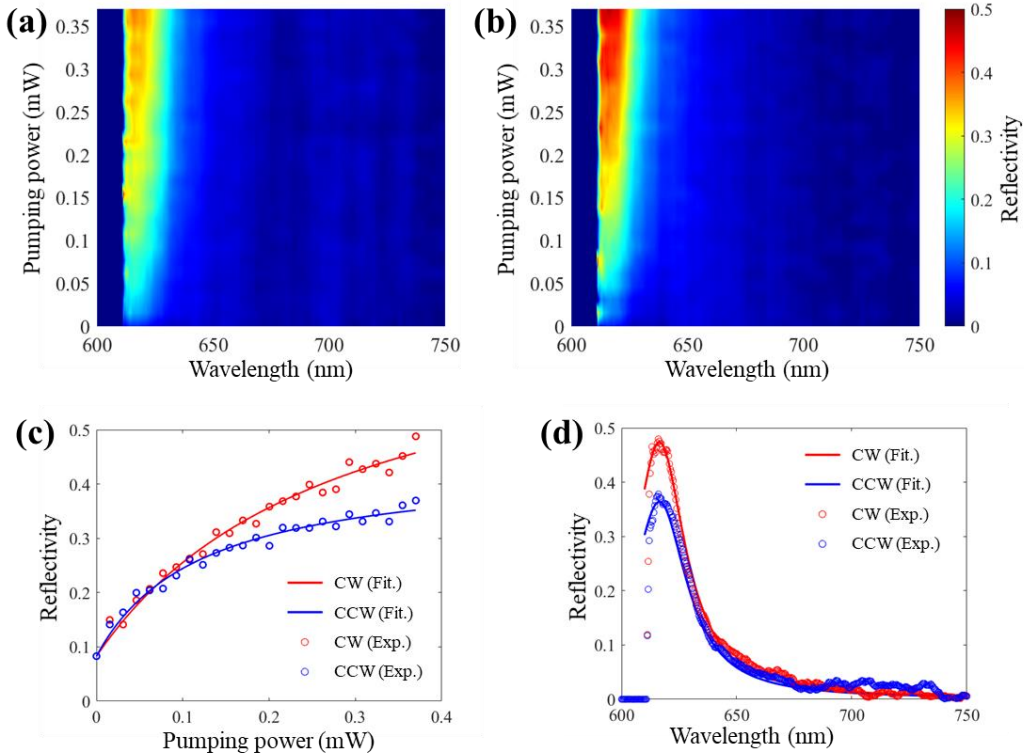

**Figure 2| Experimental demonstration of photoinduced nonreciprocal circular dichroism in WS$_2$. a,b,** Reflectivity of the WS$_2$ monolayer for probes of opposite helicities as a function of the wavelength and of the intensity of the pump with CCW helicity. **a** and **b** show the cases of CCW and CW probes, respectively. The dark blue (low signal) region on the left is due to optical filters used to eliminate the pump signal from the probe channel. **c**, Pump power dependence of the reflectivity for probes with CW and CCW helicities at $\lambda$ =616 nm, illustrating the nonreciprocal dichroic character and saturation. **d**, Wavelength dependence of reflectivity for CW and CCW probes. Solid lines show the result of fitting by a modified Fresnel equation with surface conductivity described by a Lorentzian model (see Methods for details), and the saturable absorption model described in Supplement A.

**Experimental results**

In order to experimentally verify the photoinduced dichroic response, we have performed measurements on a WS$_2$ monolayer encapsulated in hBN and transferred onto a glass substrate (hBN side on top). The image of the sample is shown in Fig. 1**c** (details of the sample preparation can be found in Methods).

The sample was pumped by 30ps circularly polarized pulses from a supercontinuum source (NKT SuperK SELECT) with 12ns pulse period at the wavelength on the blue-side and as close as possible to the exciton resonance (616 nm) to promote formation of excitons and their interactions[52]. Pumping at higher frequencies was found to lead to decreased nonreciprocal dichroic response. The intensity of the circularly polarized pump was gradually increased, and the sample reflectivity was probed with low-intensity circularly polarized beams from a halogen light source, collecting the reflected signal with a fiber-coupled spectrometer (Spectral Products SP400). The experimental results in Figs. 2**a**,**b**,**c** show that, as the pump intensity increases, the reflected probe signals with opposite projected helicities become increasingly different, confirming our hypothesis about photoinduced nonreciprocal optical dichroism. As seen from Figs. 2**a**,**b**,**d**, the dichroism shows the largest increase at the frequency of the exciton resonance (plotted separately in Fig. 2**c**), which farther proves that the mechanism responsible for nonreciprocity is associated with the difference in exciton densities at the two valleys. The clear saturable behavior of the probe signal reflection with respect to pump intensity can thus be attributed to exciton-exciton interactions (indicated by wavy lines in the schematic in Fig. 1**b**). Such interactions were previously shown to yield corrections to both the energy and the lifetime of excitons[52], giving rise to saturation in the reflectivity for pump frequencies exceeding the one of the exciton resonance.

To explain the observed photoinduced nonreciprocal dichroic behavior, we developed an analytical model based on the Fresnel equations modified by the TMD monolayer, whose optical response is described by a surface conductivity with Lorentzian dispersion[63] (see Supplement A for details). In addition, we incorporated the nonlinear effects into the Lorenz model by accounting for an increase in the exciton density as well as for the broadening of the exciton resonance due to exciton-exciton interactions[52]. We note that the spectral shift was unnoticeable in our measurements due to the already broad exciton resonance at room temperature, and hence it was neglected in our model.

The proposed model was used to fit the experimental data, enabling the retrieval of the surface conductivity tensor, with results shown in Fig. 3. Since the conductivity for each of the projected helicities is $\sigma_K = (\sigma_{xx} - \sigma_N) + \sigma_{xy}$ and $\sigma_{K'} = (\sigma_{xx} - \sigma_N) - \sigma_{xy}$, our extracted photoinduced off-diagonal (dichroic) component of the surface conductivity reaches the value $\sigma_{xy} = (\sigma_K - \sigma_{K'})/2 \approx 0.059\sigma_{xx}$ for the highest pump power at the peak reflectivity. This value of dichroic response is very large since the measurements are performed at room temperature, where intervalley scattering plays a detrimental role and, in addition, the short 30 ps pulse duration of the pump signal with 12 ns repetition rate fundamentally limits the overall nonlinear response. The fact that the dichroism does not vanish over such long integration times (compared to pump duration) indicates the presence of long-living valley-polarized excitations in the system. Indeed, while the excitons in TMDs are known to have rather short lifetimes of less than 2 ps, recent time-resolved pump-probe experimental studies have suggested that the lifetime of photoexcited free carriers can be as long as 2 ns[64] at room temperature, and may exceed values of 10 ns at cryogenic temperatures[65]. We therefore attribute the large value of the measured dichroism to the delayed relaxation of the photoexcited valley-polarized free carriers into exciton states with partial preservation of the valley-polarization. For strong pump intensities, such valley-preserving relaxation of free-carriers leads to a larger density of excitons and thus higher rate of exciton-exciton interactions at one of the valley, giving rise to an early onset of saturation at the respective valley (blue line in Fig. 2**c**). The saturation due to exciton-exciton interactions at the opposite (unpumped) valley (red line in Fig. 2**c**) has a relatively late onset and it can be solely attributed to non-valley-preserving relaxation of free-carriers into the respective valley. The schematic microscopic picture of the processes responsible for the dichroism is illustrated in Fig. 1**b**.

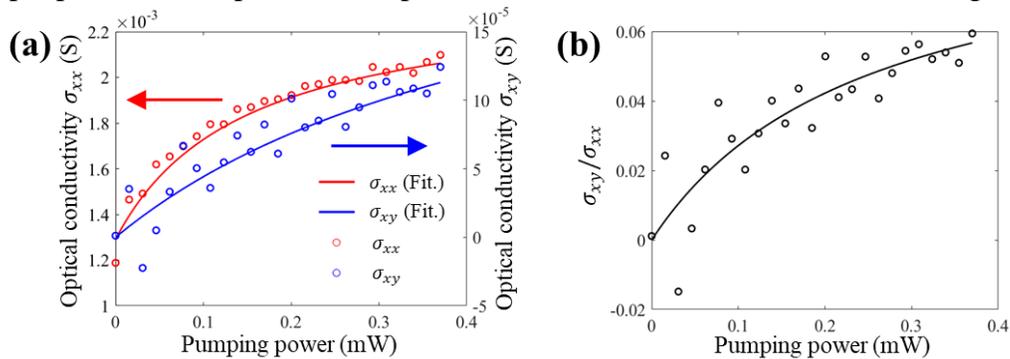

**Fig. 3| Measured dichroic surface conductivity of WS$_2$ under circularly polarized optical pump**. **a**, Diagonal element of the surface conductivity plotted alongside with the off-diagonal element. **b**, Ratio of off-diagonal and diagonal surface conductivity tensor components describing the strength of the photoinduced dichroic response. The dots in **a** and **b** show values extracted from experimentally measured reflectivities. The solid lines are the result of fitting by our analytical model (Supplement A).

**Proposed device scheme of an all-optical isolator**

To demonstrate that the photoinduced circular dichroism phenomenon can yield nonreciprocal operation, we propose a practical design of a magnet-free optical isolator relying on this effect. The proposed device is based on a silicon nitride ring resonator critically coupled to a waveguide; this scheme was recently employed for high speed modulation of light with 2D materials, graphene[66] and tungsten disulfide WS$_2$ [67]. The functionality of the device is illustrated in Figs. 4**a**, **b** and it is based on spin-orbit coupling[68] due to the non-vanishing transverse angular momentum of the evanescent optical field of guided waves[53,69]. In particular, the mode guided in the forward direction by a SiN waveguide is evanescent in the cladding, and it is characterized by CW (CCW) elliptically polarized nearfields on the left (right), as schematically shown in the inset to Fig. 4**a**. If the propagation direction of the guided wave is reversed, as in the inset to Fig. 4**b**, the handedness of the evanescent field accordingly reverses. Therefore, by placing a dichroic TMD monolayer asymmetrically with respect to the waveguide (only on the one side, as in Fig. 3**a**,**b** insets/zoom-ins) we expect different absorption rates for oppositely propagating guided waves.

Indeed, due to the dichroic response in optically pumped TMD, the different nearfield overlap of the guided wave with the surface conductivity of TMD must yield a different absorption rate for forward and backward guided modes. Such nonreciprocal absorption can be estimated using electromagnetic perturbation theory[70,71].

Taking the electric field $\boldsymbol{E}_0$ of the guided mode without the TMD monolayer as the unperturbed solution, and treating the monolayer as a perturbation, the attenuation rate due to the absorption in the TMD can be estimated to first-order to be

$$Im(\beta) = \frac{\beta_0}{\omega W_0} \int_{TMD} dS [\boldsymbol{E}_0^* Re(\hat{\sigma}_{TMD}) \boldsymbol{E}_0],$$

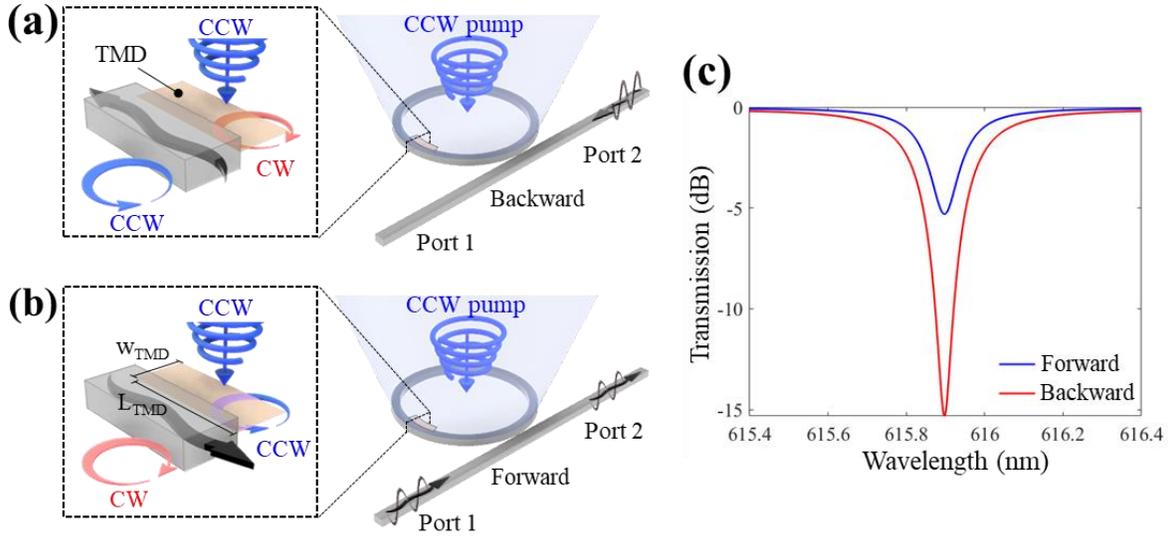

**Fig. 4| All-optical isolator device design and operation principle**. **a**, **b** SiN ring resonator loaded asymmetrically by a TMD monolayer (on the inner side of the ring only to maximize asymmetric loss), which explains nonreciprocal transmission due to different absorption of clockwise and counter-clockwise modes in the ring resonator. **c**, nonreciprocal transmission trough the waveguide with forward and backward transmission shown by blue and red lines, respectively. The ring radius $R=30$ μm, therefore the trip distance $L = 2\pi R$, t = 0.788 and **c**. $a_+ = 0.845, a_- = 0.932$, which correspond to the surface conductivity tensor $\sigma_{xx} = 2.1 \times 10^{-3}$ and $\sigma_{xy} = 1.25 \times 10^{-4}$ with parameters $\alpha = 1.79 \times 10^4$ 1/m, $\delta = 7.37 \times 10^3$ 1/m, $L_{TMD} = 6.6$ μm, the width of the TMD monolayer is $w = 50$ nm.

where $\beta_0$ is the unperturbed wavenumber of the guided wave, $S$ is the surface area, and $W_0 = 2\int_V dV[|E_0(r)|^2 \epsilon_0 \epsilon_r(r)]$ is the energy density of the unperturbed guided wave, with the integration performed over the mode volume. In this discussion, we neglect the small material loss in the unperturbed waveguide. Decomposing the evanescent field in the TMD region into the CW and CCW helicity components $\boldsymbol{E}_0 = \boldsymbol{E}_{0CW} + \boldsymbol{E}_{0CCW}$, we obtain $Im(\beta) = Im(\beta_0) + Im(\beta_K) +$

$Im(\beta_{K'})$ where $Im(\beta_0) \equiv \alpha = \frac{\beta_0}{\omega W_0} \int_{TMD} dS[|E_0|^2 \sigma_{xx}]$ and $Im(\beta_{K/K'}) = \frac{\beta_0}{\omega W_0} \int_{TMD} dS[|E_{0CW/CCW}|^2 \sigma_{K/K'}]$ are valley independent and valley polarized contributions to the attenuation caused by the absorption in the TMD. Since the pumped TMD monolayer has $\sigma_K \neq \sigma_{K'}$ and due to the evanescent field of the guided mode is chiral, we obtain a nonzero differential attenuation for forward and backward waves: $\delta \equiv Im(\beta(k > 0) - \beta(k < 0))/2 = \frac{\beta_0}{2\omega W_0} \int_{TMD} dS[(|E_{0CW}|^2 - |E_{0CCW}|^2)(\sigma_K - \sigma_{K'})]$. By applying this analysis to the modal solution for the SiN waveguide obtained with COMSOL Multiphysics, we retrieved $Im(\beta_{backford}) = 9.69 \times 10^4$ 1/m and $Im(\beta_{forward}) = 8.22 \times 10^4$ 1/m. The magnitude of such nonreciprocal attenuation, however, is not large enough to yield sufficient isolation for reasonable propagation distances. We therefore employ a resonant scheme to enhance the nonreciprocal differential absorption.

The optical isolator layout is shown in Fig. 4**a**,**b** and it consists of a SiN waveguide evanescently coupled to a SiN ring resonator. We place the circularly pumped TMD monolayer on the inner side of the ring resonator only, which ensures that the modes propagating in opposite directions have different attenuation rates. Indeed, similar to the case of the waveguide, the evanescent component of the electric field of the mode in the ring resonator carries angular momentum of opposite handedness on the inner and outer sides of the ring. The handedness of the evanescent fields flips when the propagation direction in the ring resonator reverses from CW propagating mode to CCW propagating mode, which again gives rise to a difference in absorption for the two modes.

Such different absorption rate for modes propagating in opposite directions enables to induce selectively critical coupling between the ring resonator and the waveguide[72,73] only for one propagation, yielding a strong nonreciprocal response. From here on we will use subscripts +/– to indicate CW and CCW propagation directions in the ring resonator to avoid confusion with the notations of the projected handedness of the electric field of the modes on the TMD.

According to coupled mode theory[74], the critical coupling condition for the mode propagating backward, i.e. from Port 2 to Port 1 in the SiN waveguide (Fig. 4**a**), and therefore coupling to the CW (+) mode of the ring resonator, is $t = a_+$. Here $t$ is the self-coupling of the waveguide and $a_+ = \exp[-(\alpha + \delta)L_{TMD}]$ is the round-trip loss coefficient in the ring resonator for the CW(+) mode, and $L_{TMD}$ is the length of coverage of the ring resonator by the TMD monolayer. This condition cannot be satisfied simultaneously for the waveguide mode propagating in the forward direction, i.e. from Port 1 to Port 2, thus yielding the nonreciprocal transmission. In the latter case the critical coupling condition is $t = a_-$, where $a_- = \exp[-(\alpha - \delta)L_{TMD}]$, since the guided mode now couples to the CCW (-) mode in the ring resonator which has different round-trip loss coefficient $a_- \neq a_+$.

To confirm the functionality of the proposed device, we performed CMT[74] modeling with the parameters obtained from perturbation theory using the field profiles $\boldsymbol{E}_0$ for unperturbed SiN waveguide calculated in COMSOL. For the conductivity parameters retrieved from our experimental data, we found that an isolation of 3dB can be achieved for an overall forward transmission of ~6.8%. The main limiting factor for a stronger isolation is intervalley scattering, which gives rise the non-dichroic loss in the device. Intervalley scattering can be significantly reduced by lowering the temperature. As an alternative approach, valley-polarization can be stabilized by using weak a magnetic field[75]. Here, as an example, we estimate that the effect of lowering the operating temperature would reduce the intervalley scattering fivefold (leading to approximately fivefold reduction of $\sigma_N$), in which case the nonreciprocal response in $WS_2$ can be significantly boosted. The corresponding results showing (*i*) forward transmission $S_{12}$ (from Port 1 to Port 2) and (*ii*) backward transmission $S_{21}$ (from Port 2 to Port 1) are plotted in Fig. 4**c** and clearly show strong nonreciprocal response. Note that one have to operate slightly off the exact critical coupling condition to ensure high transmission in the forward direction. Thus, for the 29.5% transmission in forward direction, the isolation of 10.0 dB can be readily achieved. We note that higher values of isolation are possible at the expense of lower backward transmission. Ideally, the performance of the proposed isolator can be farther improved with an even lower non-valley selective component of the surface conductivity $\sigma_N$, e.g., by combining lower temperature with a weak magnetic field, in which case one would operate closer to the critical coupling condition ensuring stronger isolation and higher transmission in the backward direction at the same time. Even in this case, the better compatibility and ease of integration of TMD monolayers with integrated photonic systems would render the proposed approach to nonreciprocity more practical than the use of magneto-optical materials in many applications.

**Conclusions**

To conclude, here we have experimentally demonstrated the emergence of a nonreciprocal dichroic optical response in $WS_2$ monolayer biased by circularly polarized pump field. The dichroic response was explained as the result of interaction of light with valley polarized excitons, whose interaction for higher pump intensities lead to nonlinear saturable behavior of the optical response. An analytical model of this effect was developed, which allowed to explain the experimental data and extract the surface conductivity tensor of the optically biased $WS_2$ monolayer.

The analogy of the observed dichroic response with the one of 2D electron systems in an external magnetic field suggests the possible use of optically pumped TMDs to produce magnet-free nonreciprocity. A device based on locking of transverse angular momentum of the evanescent field with the propagation direction of the guided waves was proposed. Asymmetric absorption rates due to chiral light-matter interactions with a dichroic TMD monolayer was shown to be the mechanism to achieve unidirectional critical coupling, thus producing optical isolation.

We believe that the recent progress in integration of 2D materials with existing photonic materials and devices may facilitate the impact of all-optical nonreciprocal devices based on the proposed photoinduced dichroism in photonic systems. We envision a new generation of all-optical isolators and circulators integrated into on-chip photonic systems. Moreover, the possibility to control the direction of optical isolation by simply switching the handedness of the optical pump bias makes these nonreciprocal devices switchable on the fly, enabling novel applications in classical and quantum photonic frameworks.

## Methods

### Sample fabrication:

A monolayer of $WS_2$ TMD material was exfoliated onto a thick PDMS stamp using standard tape technique and transferred to 120um thick silica substrate by home build transfer stage. The monolayer annealed at 350°C for 2 hours to remove polymer residue from the transfer process. Further, the monolayer was encapsulated with a thin hBN layer and annealed again at 350°C for another 2 hours.

### Experimental set up:

High-intensity supercontinuum light-source SuperK Extreme with connected LLTF Contrast tunable high-resolution bandpass filter generated light beam with 2nm bandwidth and the tunable wavelength in the range 0.4-1.0 μm. The sample was pumped with supercontinuum pulsed laser of 30ps pulse width and 12ns pulse period and wavelength of light was set to 600 nm close to exciton resonance (616 nm) at room temperature. The polarization of the excitation beam was set to circular polarization by using a combination of linear polarizer and quarter wave plate. A 50X (Olympus) long working distance microscopic objective was used to excite the monolayer $WS_2$ with 1um spot size at $30^0$ angle of incidence. A low intensity white light beam from halogen light source was used as probe beam with circular polarization set up by another set of linear polarizer and quarter wave plate. A long working distance 50X microscopic objective (Boli optics) with 1um spot size at the focus was used to probe the sample reflection spectrum. The reflected probe beam from the sample was passed through 610 nm long pass filter to cut of the excitation beam and analyzed the spectrum by using fiber coupled spectrometer (Spectral Product 400).

### Data availability

Data that are not already included in the paper and/or in the Supplementary Information are available on request from the authors.

### Acknowledgements

The work was supported by the National Science Foundation with grants No. DMR-1809915, EFRI-1641069, by the Defense Advanced Research Project Agency Nascent Program, and by the Simons Foundation.

## Author contributions

All authors contributed extensively to the work presented in this paper. SG and YK contributed equally to this work.

## Author Information

# Supplementary information

# All-optical nonreciprocity due to valley polarization in transition metal dichalcogenides

Yuma Kawaguchi[1], Sriram Guddala[1], Kai Chen[1], Andrea Alù[3,2,1], Vinod Menon[2,4], and Alexander B. Khanikaev[1,2]

[1]Department of Electrical Engineering, Grove School of Engineering, City College of the City University of New York, 140th Street and Convent Avenue, New York, NY 10031, USA.
[2]Physics Program, Graduate Center of the City University of New York, New York, NY 10016, USA.
[3]Photonics Initiative, Advanced Science Research Center, City University of New York, New York, NY 10031, USA
[4]Department of Physics, City College of New York, 160 Convent Ave., New York, NY 10031, USA

# Supplement A

## A1. Analytical models to fit the experimental results

The reflectivity spectra for CW and CCW probe at maximum pump power (Figure.2**d**) was fitted by the Fresnel equation modified by sheet conductivity[1],

$$|r|^2 = \left|\frac{n_2-n_1-Z_0\tilde{\sigma}}{n_2+n_1-Z_0\tilde{\sigma}}\right|^2. \quad (A1)$$

For the CW and CCW probe reflectivities, one considers the respective sheet conductivities $\tilde{\sigma}_{CW} = \sigma_{xx} + \sigma_{xy}$, $\tilde{\sigma}_{CCW} = \sigma_{xx} - \sigma_{xy}$, respectively, which yield

$$|r_{CW}|^2 = \left|\frac{n_2-n_1-Z_0(\sigma_{xx}+\sigma_{xy})}{n_2+n_1-Z_0(\sigma_{xx}+\sigma_{xy})}\right|^2, \quad (A2)$$

$$|r_{CCW}|^2 = \left|\frac{n_2-n_1-Z_0(\sigma_{xx}-\sigma_{xy})}{n_2+n_1-Z_0(\sigma_{xx}-\sigma_{xy})}\right|^2, \quad (A3)$$

where $Z_0 = \sqrt{\mu_0/\epsilon_0}$ is the free space impedance, and $n_1, n_2$ are the respective refractive indices of the superstrate and substrate surrounding the TMDs monolayer. The off-diagonal component $\sigma_{xy}$ gives rise to the difference in reflectivities of the CW and CCW probes.

The optical permittivity and sheet conductivity of TMD monolayer, which are the functions of the photon energy $E$, were fitted by Lorentzian dispersion model[2],

$$\epsilon(E) = 1 + \frac{f}{E_{ex}^2+E^2-iE\gamma}, \quad (A4)$$

$$\tilde{\sigma}(E) = -\frac{i\epsilon_0 E d}{h}[\epsilon(E)-1] = -\frac{i\epsilon_0 E d}{h}\frac{f}{E_{ex}^2+E^2-iE\gamma}, \quad (A5)$$

where $f$ is the oscillator strength, $E_{ex}$ is the exciton resonance energy and $\gamma$ is the linewidth of the exciton resonance, and all represent pump intensity dependent parameters, while $d$ is the thickness of TMD monolayer and $h$ is the Planck constant. The respective parameters for the case of maximal pump power are given in Table A1.

**Table A1|** Fitting parameters at $I_{CCW} = 0.38\ (mW)$ of Lorentzian type model in Fig.**2d**

| $f$ (eV$^2$) | $E_{ex}$ (eV) | $\gamma$ (eV) | $d$ (nm) |
|---|---|---|---|
| 236 | 2.018 | 0.15 | 0.618 |

Assuming only linear corrections to the linewidth and oscillator strength due to the pump, i.e. $\gamma = \gamma_0 + \gamma_1 I_{CCW}$ and $f = f_0 + f_1 I_{CCW}$, and taking into account the lack of dependence of rhe resonance position observed in the experiment, we obtain an approximate expression for the intensity dependent conductivity,

$$\sigma_{xx,xy} = \frac{AI_{CCW}}{1+\alpha I_{CCW}} + \sigma_0. \tag{A6}$$

The expression (A6) has the standard form of saturable absorption and has straightforward physical interpretation with the parameter $A$ being the rate of change of exciton density due to the pump, and the parameter $\alpha$ reflecting the spectral broadening of the resonance due to the exciton-exciton interactions. Figure **3a** of the main text shows fitting results for the optical sheet conductivities $\sigma_{xx}$ and $\sigma_{xy}$ using equation (A6), while Table A2 gives the corresponding values of fitting parameters at the exciton resonance. The term $\sigma_0$ in (A6) corresponds to the optical conductivity without pumping and it is zero for $\sigma_{xy}$.

**Table A2|** Fitting parameters of optical sheet conductivities

|  | $A$ | $\alpha$ | $\sigma_0$ |
|---|---|---|---|
| $\sigma_{xx}$ | 0.0073 | 6.83 | 0.0013 |
| $\sigma_{xy}$ | 0.00053 | 1.99 | 0 |

By combining the expressions (A1-A3) and (A6) we derive the reflectivities, which are saturable with the pumping intensity $I_{CCW}$. For the simplest case of no substrate $n_{1,2} = 1$, and by considering the negligible value of $\sigma_0 \approx 0$, after all the substitutions we obtain the following expression:

$$|r|^2 = \left|\frac{Z_0\left(\frac{AI_{CCW}}{1+\alpha I_{CCW}}\right)}{2-Z_0\left(\frac{AI_{CCW}}{1+\alpha I_{CCW}}\right)}\right|^2 = \left|\frac{Z_0 AI_{CCW}}{2(1+\alpha I_{CCW})-Z_0 AI_{CCW}}\right|^2$$

$$= \left|\frac{(Z_0 AI_{CCW})^2}{4(1+\alpha I_{CCW})^2 - 4(1+\alpha I_{CCW})(Z_0 AI_{CCW})+(Z_0 AI_{CCW})^2}\right|$$

$$= \left|\frac{(Z_0 AI_{CCW})^2}{4+(8\alpha-4Z_0 A)I_{ccw}+(4\alpha^2-4\alpha Z_0 A+Z_0^2 A^2)I_{CCW}^2}\right|. \tag{A7}$$

By adding a background correction due to the substrate, we arrive at the following expression describing the saturable behavior for the probes with opposite helicities

$$|r_{CW,CCW}|^2_{max} = \frac{A_{CW,CCW}I_{CCW}}{\sqrt{4+\alpha_{CW,CCW}I_{CCW}+\beta_{CW,CCW}I^2_{CCW}}} + |r_0|^2, \quad (A8)$$

where $|r_0|^2$ is the reflectivity without CCW pumping $I_{CCW} = 0$. The experimentally obtained reflectivity spectra for the CW and CCW probes in Fig. 2**c** of main text were fitted with Eq. (A8), and the respective fitting parameters are given in Table A3.

**Table A3**| Fitting parameters of the saturable absorption

| $r_0$ | $A_{CW}$ | $\alpha_{CW}$ | $\beta_{CW}$ | $A_{CCW}$ | $\alpha_{CCW}$ | $\beta_{CCW}$ |
|---|---|---|---|---|---|---|
| 0.0828 | 6.780 | 112.6 | -21.86 | 11.97 | 464.1 | 597.8 |

## A2. Perturbation theory for the waveguide interacting with the dichroic 2D material

The simulation of the waveguide was performed in COMSOL Multiphysics. Experimentally obtained optical conductivities were then used to obtain the imaginary parts of the propagation constant in the waveguide $\beta$, which describes the attenuation of the guided wave, and is given by the expression:

$$\mathrm{Im}(\beta) = \frac{\beta_0}{\omega W_0} \int_{TMD} dS [\mathbf{E}_0^* Re(\hat{\sigma}_{TMD}) \mathbf{E}_0], \quad (A9)$$

where

$$W_0 = 2\int_V dV [|E_0(r)|^2 \epsilon_0 \epsilon_r(r)], \quad (A10)$$

and $Re(\hat{\sigma}_{TMD})$ is the real part of the optical conductivity of the TMDs monolayer, $\mathbf{E}_0$ is the absolute value of the electric field, and $W_0$ is the energy density of the guided wave (the core plus the cladding). The imaginary part of the propagation constant $\alpha$ due to the diagonal component of conductivity tensor, i.e. non-discriminating handedness of the field helicity, is then given by

$$\alpha = \frac{\beta_0}{\omega W_0} \int_{TMD} dS [|E_0|^2 \sigma_{xx}], \quad (A11)$$

while the helicity dependent absorption differential is defined by the off-diagonal component of conductivity as

$$\delta = \frac{\beta_0}{2\omega W_0} \int_{TMD} dS [|E_{0CW}|^2 - |E_{0CW}|^2)(\sigma_K - \sigma_{K'})]. \quad (A12)$$

Since the conductivity is uniform across the TMD monolayer, the respective terms can be moved out from the integrals and will appear as factors. Then, the integrals defining the nonreciprocal absorption are evaluated from numerically calculated field profiles with values given in Table A4.

**Table A4|** Values of the perturbation theory integrals used for the evaluation of the propagation constant

| Expression | Evaluated integration value |
|---|---|
| $\int_{TMD} dS |E_{0CW}|^2$ | 189.1 V$^2$ |
| $\int_{TMD} dS |E_{0CCW}|^2$ | 1029.3 V$^2$ |
| $\int_{core} dV |E_0(r)|^2$ | 0.000419 V$^2$m |
| $\int_{clad} dV |E_0(r)|^2$ | 0.00111 V$^2$m |
| $\sigma_{xx}$ | 0.0021 S |
| $\sigma_{xy}$ | 0.00125 S |

# Supplement B

## B1. Photoinduced nonreciprocal circular dichroism in WS$_2$ for CW pump

The photoinduced nonreciprocal circular dichroism in WS$_2$ monolayer measured for the CW pump is shown in Fig. S1. It can be noticed that in this case CW probe reflectivity shows the saturation, whereas for the case of CCW pump of the main text, it is the CCW probe reflectivity shows saturation (Fig. 2 of the main text).

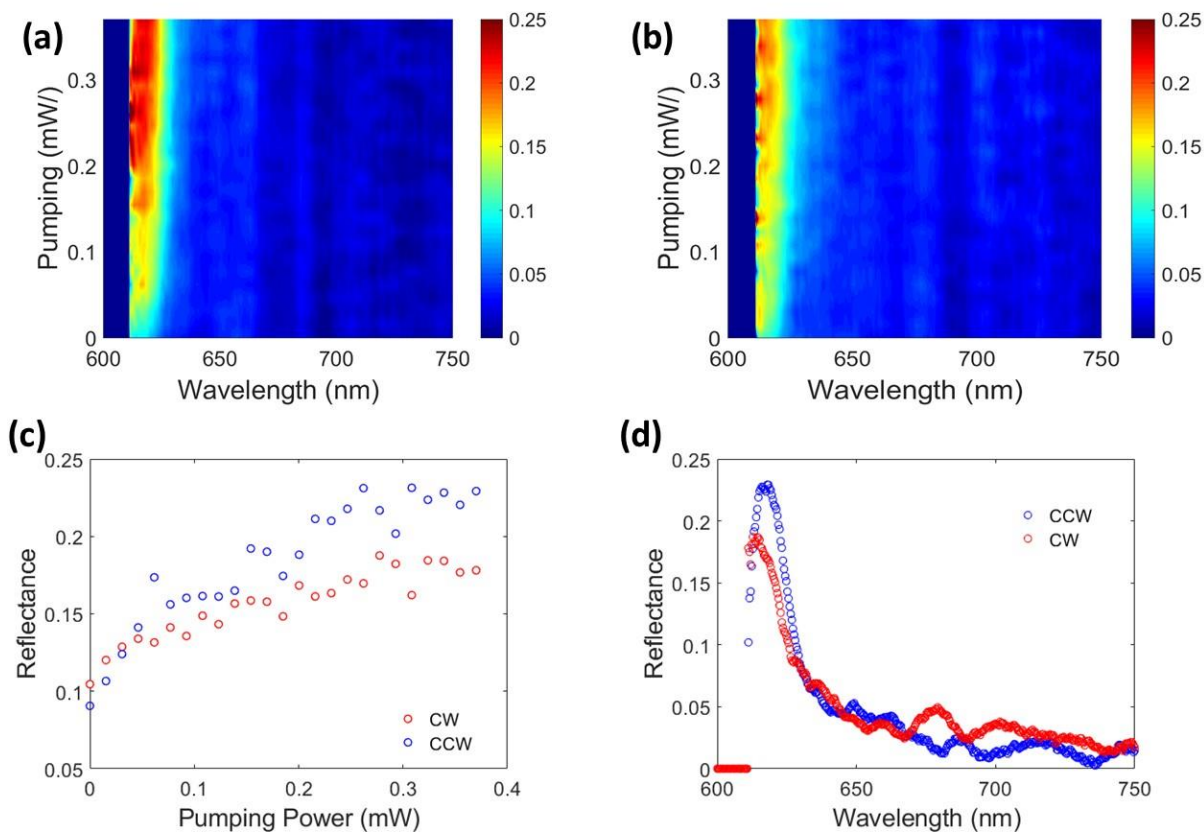

**Figure S1| Experimental demonstration of photoinduced nonreciprocal circular dichroism in WS$_2$ for the CW pump. a,b**, Reflectivity of WS$_2$ monolayer with respect to probe of two opposite helicities as the function of the wavelength and of the intensity of the pump with CW helicity. **a** and **b** show the cases of CCW and CW probes, respectively. The dark blue (low signal) region on the left is due to optical filter used to eliminate the pump signal from the probe channel. **c**, pump power dependence of the reflectivity of the probes with CW and CCW helicities at $\lambda$ =616 nm illustrating dichroic character and saturation. **d**, wavelength dependence of reflectivity for the CW and CCW probes.

## B2. Schematic of experimental setup

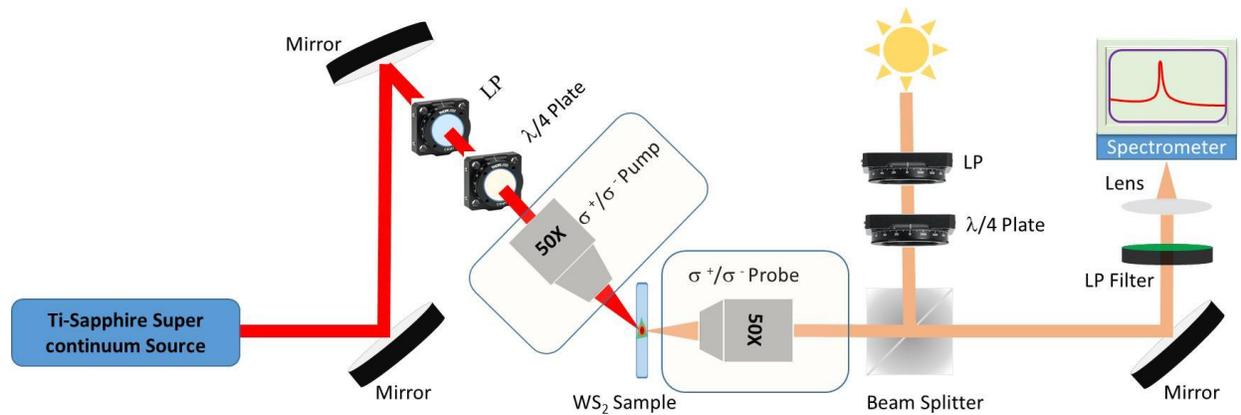

LP: linear polarizer; LP Filter: Long pass filter.

**References**

1. Stauber, T., Peres, N. M. R. & Geim, A. K. Optical conductivity of graphene in the visible region of the spectrum. *Phys. Rev. B - Condens. Matter Mater. Phys.* **78**, 085432 (2008).

2. Li, Y. *et al.* Measurement of the optical dielectric function of monolayer transition-metal dichalcogenides : *Phys. Rev. B* **205422**, 1–6 (2014).